\documentclass[onecolumn,amsmath,amssymb, 12pt]{iopart}
\usepackage{graphicx}
\usepackage{dcolumn}
\usepackage{bm}
\usepackage[tight]{subfigure}
\begin{document}

\title{{\Large $J/\psi$ production in pp collisions at $\sqrt{{\rm s}}$=200GeV in the STAR experiment }}

\author{\footnotesize MAURO R. COSENTINO for the STAR COLLABORATION}

\address{Instituto de F\'{i}sica, Universidade de S\~ao Paulo, Rua do Mat\~ao - travessa R, 187 \\  S\~ao Paulo, SP 05508-090,Brazil}

\ead{mcosent@dfn.if.usp.br}

\hspace{1.6cm}{\it \footnotesize  Proceedings for the Quark Matter 2008}

\begin{abstract}
\leftskip1.0cm
\rightskip1.0cm
In this work we present the $J/\psi$ measurement in p+p collisions within the STAR collaboration Quarkonium program. This measurements aim to be the baseline measurement of a more comprehensive systematic study of quarkonium states production in order to understand their in medium modification. Here we report the total cross section and $p_T$ distribution, and find them to be consistent with pQCD CEM predictions as well as to previous measurements at the same center-of-mass energy.
\end{abstract}


\section{Motivation}

In relativistic heavy ion collisions it is expected to occur the formation of a hot and dense matter where quarks and gluons would move freely, experiencing hadronic deconfinement. One of the several experimental signatures of such state of nuclear matter is the suppression of the $J/\psi$ production by color screening, as firstly predicted by Matsui and Satz \cite{MatsuiSatz}. Recent lattice QCD calculations predict that the temperature achieved by the medium is not high enough to dissolve the promptly formed $J/\psi$'s, but would be enough to melt higher energies $c\bar{c}$ states, resulting in a suppression pattern due to feeddown. In order to verify such suppression, there must be a baseline measurement to compare with the $J/\psi$ production in A + A collisions. This baseline is the purpose of this work. The STAR experiment developed a specific trigger for $J/\psi$ measurement in p +p collisions for the RHIC run of 2006, and the results of this measurement is presented in this work.

\section{Experimental Setup}

STAR detector~\cite{star} is a multipurpose experiment composed of several subsystems suited to measure many different observables in the central rapidity region. Its large acceptance ($\vert\eta\vert<$1 and 0$<\phi<$2$\pi$)  assures STAR the capability of measure many of the quarkonium states, including the $J/\psi$ and $\Upsilon$ family, through their $e^+e^-$ decay channel. The subsystems used to do it are the Time Projection Chamber (TPC), the Barrel Electromagnetic Calorimeter (BEMC) and, for $J/\psi$, the Central Trigger Barrel (CTB). Also, a specific trigger was designed for this measurement.

The TPC is the main STAR subsystem. It is designed to reconstruct the tracks of charged particles, giving precise information on momentum and ionization energy loss ($dE/dx$). With the TPC it is possible to make a first set of particle identification cuts for electrons (in this text every mention to electrons means the same for positrons).

The BEMC is a sample calorimeter with full azimutal coverage, surrounding the TPC, and is divided in 120 modules of 40 towers each (20 in $\eta$, 2 in $\phi$.). Each module range from $\eta=$0 to 1 or -1 and have azimutal coverage of 2$\pi$/60. The towers are stacks of 21 alternating layers of scintilator and lead absorber, corresponding to $\sim$21$X_0$ and with geometrical acceptance of $\Delta\eta\times\Delta\phi=$0.05$\times$0.05 each. The energy resolution in the towers is $\frac{dE}{E}\sim\frac{16\%}{\sqrt{E}}$. The first 2 lead-scintilator layers compose the Pre Shower Detector (PSD) and at the position of 5$X_0$ from the beamline, sits the Shower Maximum Detector (SMD). For this work only the BEMC towers were used.

Finally, the CTB is a detector made of 120 trays containing 2 scintilator slats each.It is located between the TPC and the BEMC, with the same acceptance coverage of the last one. The CTB was originally developed to trigger events based on the detection of charged particles passing through it, but in this work it was used as an auxiliary system to trigger $J/\psi$ events. Each one of the scintilator slats matches the same area of $\sim$20 BEMC towers.

\subsection{Trigger}

In order to accomplish the measurement of a rare probe such as the $J/\psi$ we needed to develop a specific event trigger setup. This trigger setup was divided in two levels, the first of them (L0) was a topologic trigger designed in the hardware level. In L0 the setup divides the BEMC logically in 6 sections in $\phi$, and make a requirement that at least two BEMC towers have energy readings above the threshold of 1.2 GeV. The sectors where these towers are must not be adjacent. Once these requirements are met, L2 starts.

The L2 is a software algorithm that takes the input from the L0 to make the final decision on the event. First of all, the CTB information is used to veto towers from L0 that received their energy from incident photons. This is done by requiring an ADC signal in the CTB slat that corresponds geometrically to the chosen tower, ensuring that the signal is from a charged particle as the CTB slats are not sensitive to photons. The non vetoed L0 towers are then used as seeds to create clusters of 3 towers: the seed plus the two adjacent towers with the highest ADC readings. With the clusters created then their position is computed as the weighed average of the three towers position, using their ADC readings as weight. Once the positions are computed, the clusters are grouped into pairs and a straight line from the clusters to the vertex position (in this case 0,0,0) and the angle $\theta$ between these two lines is obtained. With the information of the clusters energy and the angle between them, then a approximated invariant mass is calculated by
\begin{equation}
 M = \sqrt{2E_iE_j\left(1-\cos{\theta_{ij}}\right)}
\end{equation} 
where $E_{i(j)}$ is the energy of the cluster $i(j)$ and $\theta_{ij}$ is the angle between them. This expression is just a reasonable approximation since the $e^+e^-$ tracks are not straight lines. If 2.2$<M<$5.0 GeV/c$^2$ for at least one of the possible pairs of clusters, then the event is recorded.

The time to accomplish L0 is $\leq$1 $\mu$s and the decision at the L2 is taken up to 500 $\mu$s.

\section{Analysis and discussion}

The first important part of the analysis is to cleanly and efficiently identify electrons and positrons, since they are the observed decay channel in this work. The identification begins with a good track selection criteria, which are based on a mininum number of ionization points of each track in the TPC gas, in order to assure a more precise reconstruction of momentum and $dE/dx$ for each track. Once the good tracks are selected, then the $dE/dx$ information for each track is taken into account. Tracks with $dE/dx$ values consistent with the expected value for electrons and significantly different values from the expected values of hadrons are selected. These selected tracks, candidates to electrons, are then extrapolated to the BEMC towers, and the energy of the corresponding tower, $E_{tow}$ is taken to compute the ratio with its corresponding track momentum, $p_{track}/E_{tow}$. Considering the ultra-relativistic state of the electrons and the fact that they must loose all of their energy on the tower one should expect $p_{track}/E_{tow}$=1. So the last identification cut is to selected the candidates with $p_{track}/E_{tow}\sim$1.

  \begin{figure}
    \begin{center}
    \subfigure[Net signal and simulated line shape.]
      {
      \label{fig:jpsiLineShape}
      \includegraphics[width=0.47\textwidth]{{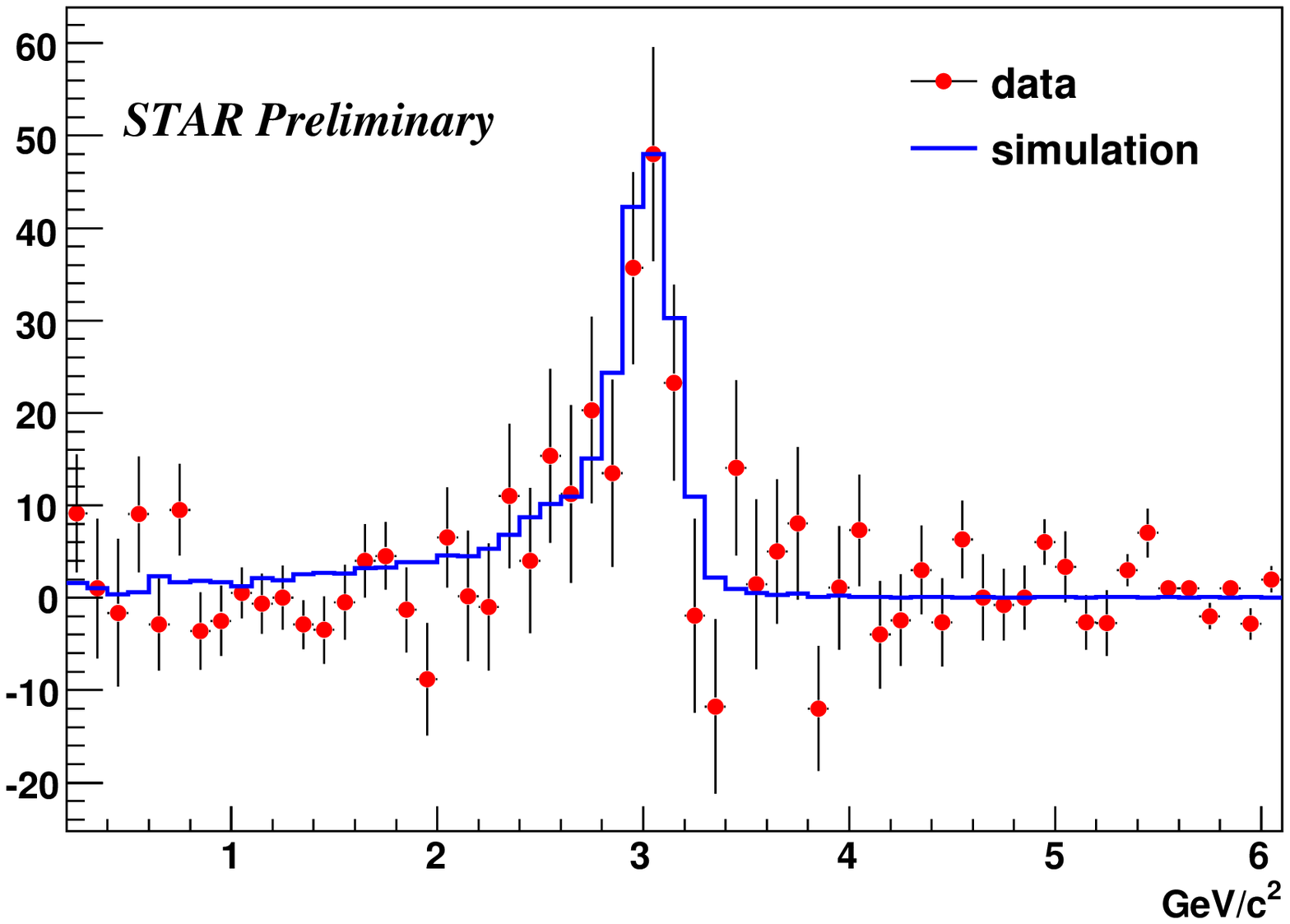}}
      }
    \subfigure[$J/\psi$ $p_T$ spectrum and pQCD CEM curves.~\cite{ramona}]
      {
      \label{fig:jpsiPt}
      \includegraphics[width=0.47\textwidth]{{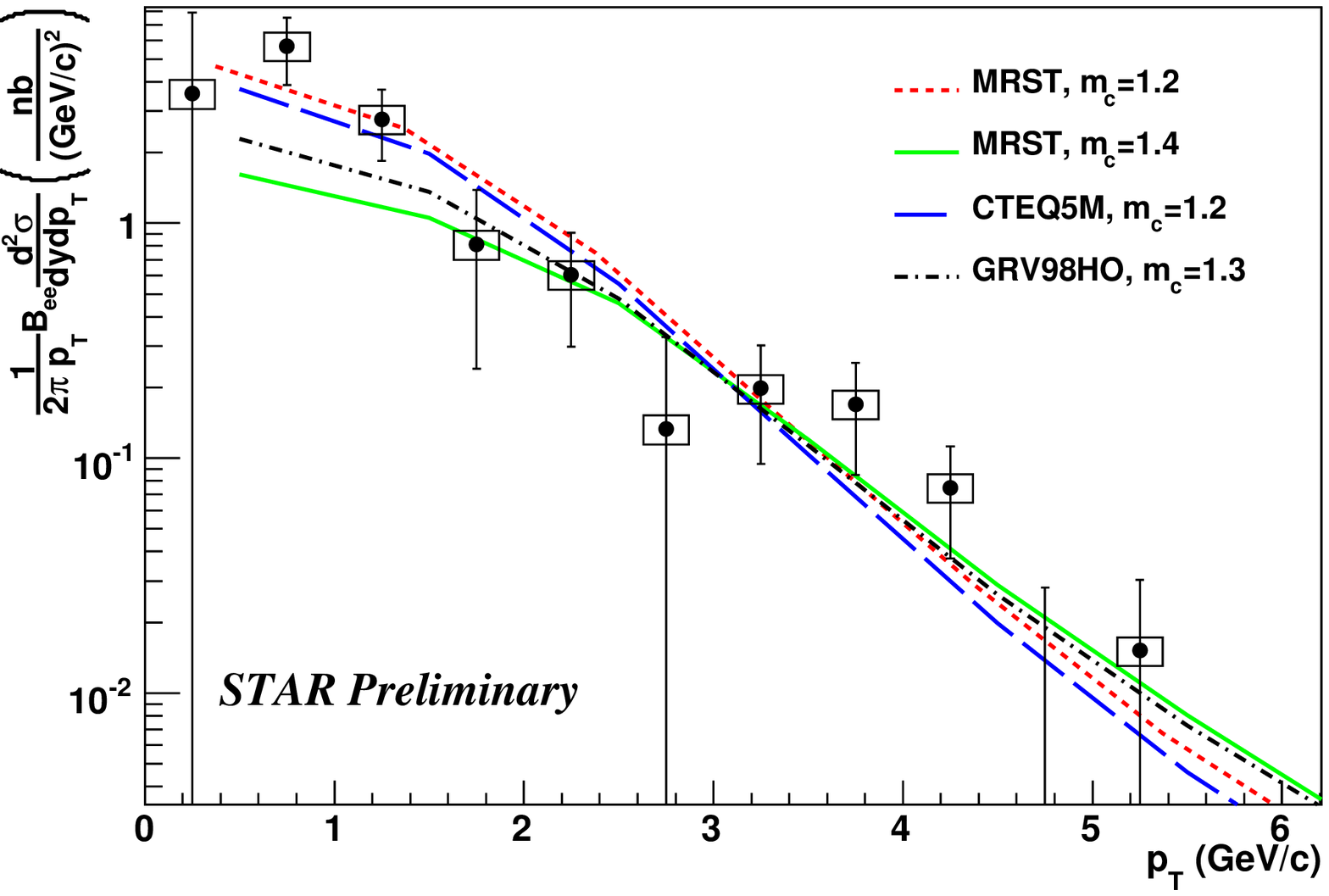}}
      }
    \end{center}
    \caption{Invariant mass spectra from di-electron pairs and $p_T$ spectrum.}
    \label{fig:mass}
    \end{figure}  

Once the electrons are selected for a given event, they are combined into pairs and have their invariant mass computed by the expression
\begin{equation}
 M = \sqrt{2p_1p_2\left(1-\cos{\theta_{12}}\right)}
\end{equation} 
where $p_i$ is the magnitude of the 3-momentum of the $i$-th particle and $\theta_{12}$ the angle between them in the point of closest approximation of their tracks. The pairs with unlike signed charges fill the signal mass spectrum, while the like signed charges contribute to estimate the background. The background spectrum is computed by twice the geometric mean of the $e^+e^+$ and $e^-e^-$ mass spectra.The net signal is then the subtraction between signal and background, and the significance of the signal, found to be $S=$ 5.9 $\sigma$, is calculated by
\begin{equation}
 S = \frac{T-B}{T+B}
\end{equation} 
where $T$ is the yield of $e^+e^-$ and $B$ is the yield of the background as described above. Figure \ref{fig:jpsiLineShape} presents the net signal mass spectrum superimposed to a GEANT~\cite{Geant3} simulation. From this figure is possible to see the good agreement of the simulated $J/\psi$ peak shape, describing very well the width and position of the experimental data. The long tail to the left due to the bremsstrahlung effect on the di-electron pairs, as described by the simulation, is also visible in the data spectrum.The yield was calculated integrating the data in the range 2.2$<M<$3.2 GeV/c$^2$, and the computed value was 207$\pm$35 $J/\psi$'s.
Considering all the efficiency corrections, the differential cross section obtained was
\begin{equation}
\label{eq:cross-sec}
 \mathcal{B}_{ee}\times\frac{d\sigma}{dy}\vert_{y=0} = 57\pm10\left(stat\right)\pm9\left(syst\right) {\rm nb}
\end{equation} 
which is consistent with previous measurement at the same $\sqrt{s}$~\cite{jpsiPhenix}. $\mathcal{B}_{ee}$ ($=$ 5.93\%) is the branching ratio in the $e^+e^-$ decay channelThe efficiency errors were calculated according to~\cite{effError}.

\begin{figure}
    \begin{center}
    \subfigure[$\langle p_T^2\rangle$ and world data~\cite{wdata} as function of $\sqrt{s}$.]
      {
      \label{fig:jpsiPt2}
      \includegraphics[width=0.47\textwidth]{{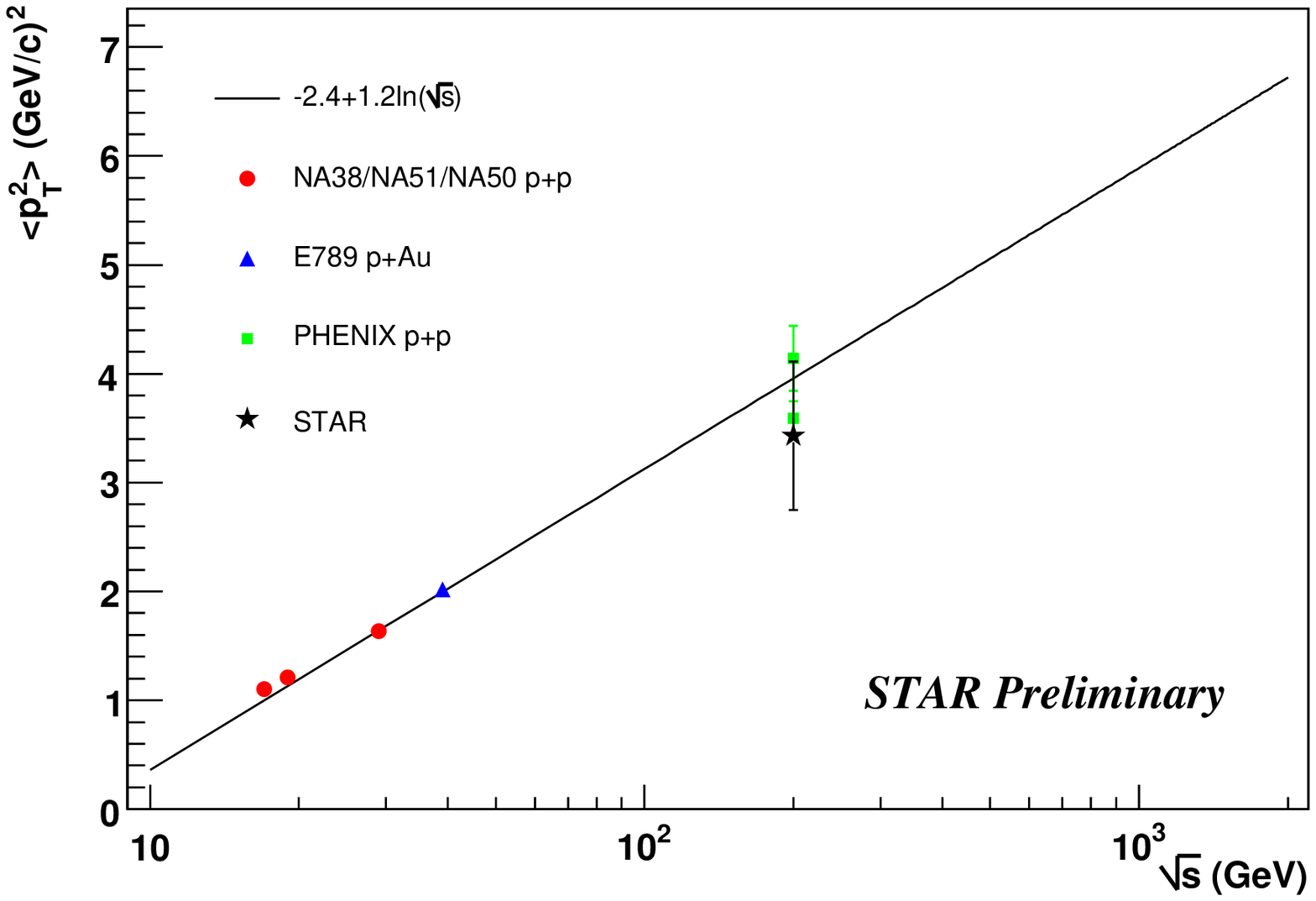}}
      }
    \subfigure[$\sigma^{incl}_{tot}$ and world data~\cite{wdata} as function of $\sqrt{s}$.]
      {
      \label{fig:jpsiSecChoqE}
      \includegraphics[width=0.47\textwidth]{{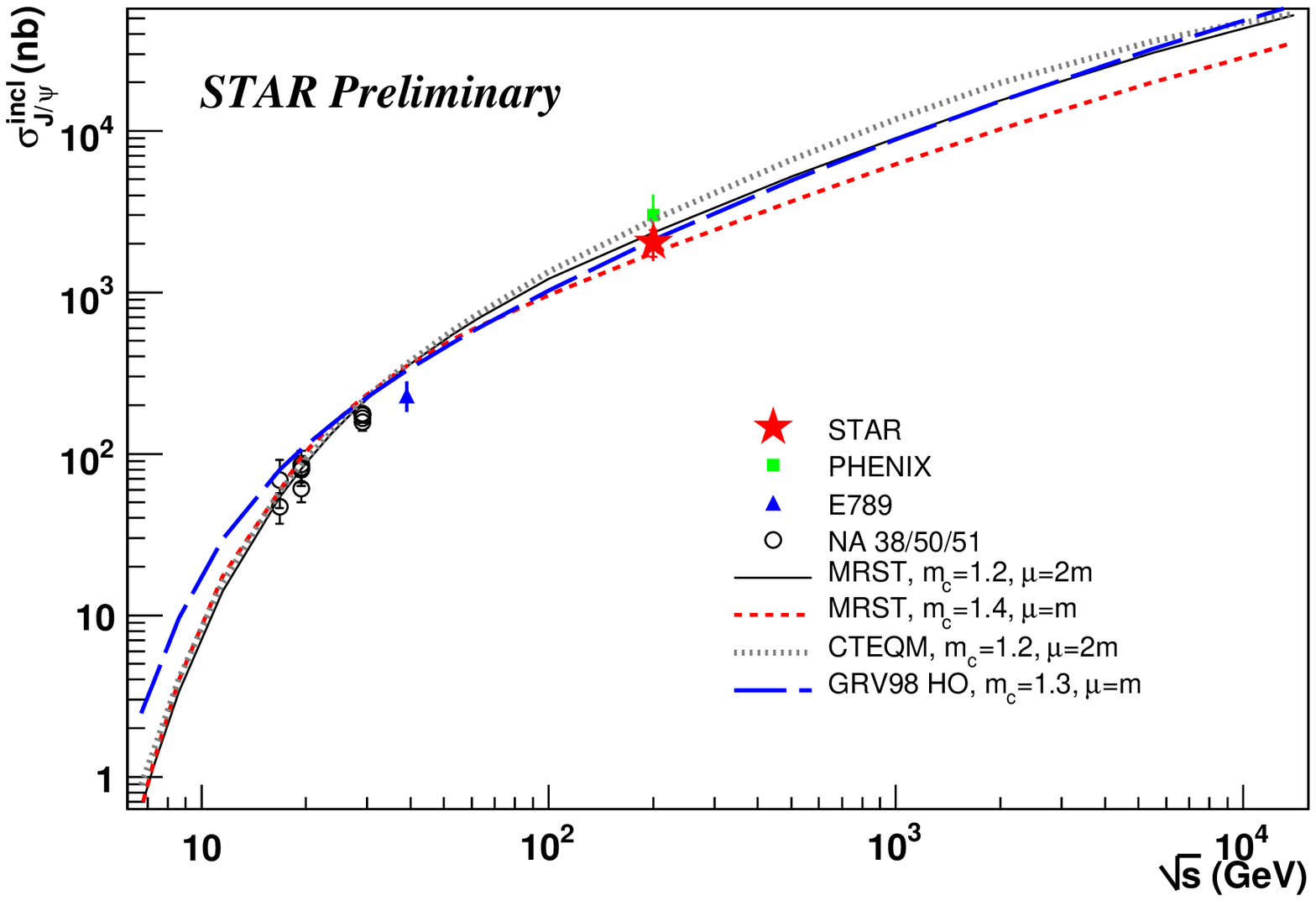}}
      }
    \end{center}
    \caption{$\sqrt{s}$ systematics for $\langle p_T^2\rangle$ and $\sigma^{incl}_{tot}$.}
    \label{fig:ptSpectra}
    \end{figure}

With the available statistics was also possible to construct the $p_T$ spectrum, presented in figure \ref{fig:jpsiPt}, and derive the $\langle p_T^2\rangle$ from it. The value obtained for our measurement is $\langle p_T^2\rangle=$3.43$\pm$0.68. \ref{fig:jpsiPt} also presents pQCD-CEM calculations~\cite{ramona} that are fairly consistent with our data. In \ref{fig:jpsiPt2} it is possible to see the consistency of our measurement with PHENIX~\cite{jpsiPhenix} and with other experiments~\cite{wdata} systematics as a function of $\sqrt{s}$. In addition to this, the total inclusive cross-section as function of the center-of-mass energy is presented in figure \ref{fig:jpsiSecChoqE}, corroborating the consistency of our data with pQCD-CEM calculations~\cite{bejidian} and world data. The value for the total inclusive cross section was obtained making an extrapolation of the value at mid-rapidity (y=0) multiplied by the mean ratio, $\langle\sigma^{incl}/\sigma^{y=0}\rangle$, which is the average of the same ratio for the each one of the four theoretical calculations of~\cite{ramona}.

\section{Summary}

In this work we presented the preliminary results of the first $J/\psi$ measurement of the STAR collaboration. Together with the $\Upsilon$ measurement in p+p collisions~\cite{upsilon}, this is the first successful result of the STAR Quarkonium program. The results obtained with this work showed to be consistent with previous measurements at this same energy, as well as with theoretical predictions of pQCD-CEM model. The $\langle p_T^2\rangle$ and $\sigma_{tot}^{incl}$ are consistent with PHENIX measurement and with world data trends as functions of $\sqrt{s}$.

This work was supported by Brookhaven National Laboratory and the Brazilian support agencies CNPq and CAPES.

\end{document}